\documentclass{PoS}

\title{Layer World: Living on a layer in 5D SU(3)}

\ShortTitle{Layers in 5D SU(3)}

\author{\speaker{K.~Petrov}\\
       Laboratoire de Physique Theorique, Universite Paris-sud, Orsay\\
        E-mail: \email{Konstantin.Petrov@th.u-psud.fr}}

\abstract{We suggest another approach to five-dimensional non-isotropic gauge
  theory. Using non-perturbative technique we show that already modest
  interaction anisotropy confines heavy bound states to four-dimensional
  layers, while free quarks propagate with almost no penalty into the fifth dimension.}

\FullConference{The XXVII International Symposium on Lattice Field Theory - LAT2009\\
		 July 26-31 2009\\
		 Peking University, Beijing, China}

\begin{document}

\section{Introduction}
The question of dimensionality of our world is closed and re-opened with
clockwork regularity. Relative silence since the days of Kaluza and Klein
turned into a roar with the advent of string theory and competing field
theories. Since then string theorists are trying to decrease number of
dimensions to come to four, and field theorists are trying to increase them
to explain all particles. Many string and all field theories of this kind are
a reiteration of the Kaluza-Klein mantra - ``compact external dimensions lead to
new massive particles''. None of these theories have been successful so far. 

The reason for that may be a certain lack of attention, as the extra-dimensional gauge theories are believed to be
non-renormalizable, which makes them somewhat exotic and therefore less
studied.  However, to our best knowledge, such non-renormalizability for the lattice regularization has not been proven.

Yet, there is a way out of this contradiction. We accepted already that time
is different from the other dimensions, and though we managed to formulate the
four dimensional field theory via an effective euclidian theory, it still
remains different. So without much effort we may imagine that the coupling in 
higher dimensions is either weaker or stronger than in four. Here we consider
the former possibility. This changes the rules of the game  and gives us more
possibilities for renormalization at the cost of extra parameter.

It was pioneered in the classical work of Nielsen and Fu \cite{Fu:1983ei},
who, after destroying  possibility of having a discrete field theory in four
dimensions went on to study higher dimensional theories. Their hope was to
find a high-dimensional space which would, at  certain values of couplings
turn into weakly or non-interacting layers of lesser dimension. Let us see how
that worked out and how we can extend and reinterpret their findings using
modern warfare of lattice simulations \cite{chroma}.

      %%%%%%%%%%%%%%%%%%%%%%%%%%%%%%%%%%%%%%%%%%%%%%%%%%%%%%%%%%%%%%%%%%%%%%%%%%%%%%%%%%%%%%%%%%%%%%%%%%%%%%%%%%%%
      
\section{Brave New World}
       
Let us assume that the world is either fundamentally or at least effectively five-dimensional and its strong dynamics is determined, in the first approximation,
by a simple gauge theory. Furthermore, let us assume that the coupling in the fifth
direction is different from the couplings in other directions. It has an obvious
disadvantage  of introducing 
an extra parameter, but so do Kaluza-Klein-like theories.  The action, in
terms of plaquette variables then reads:

        \includegraphics[width=0.7\linewidth]{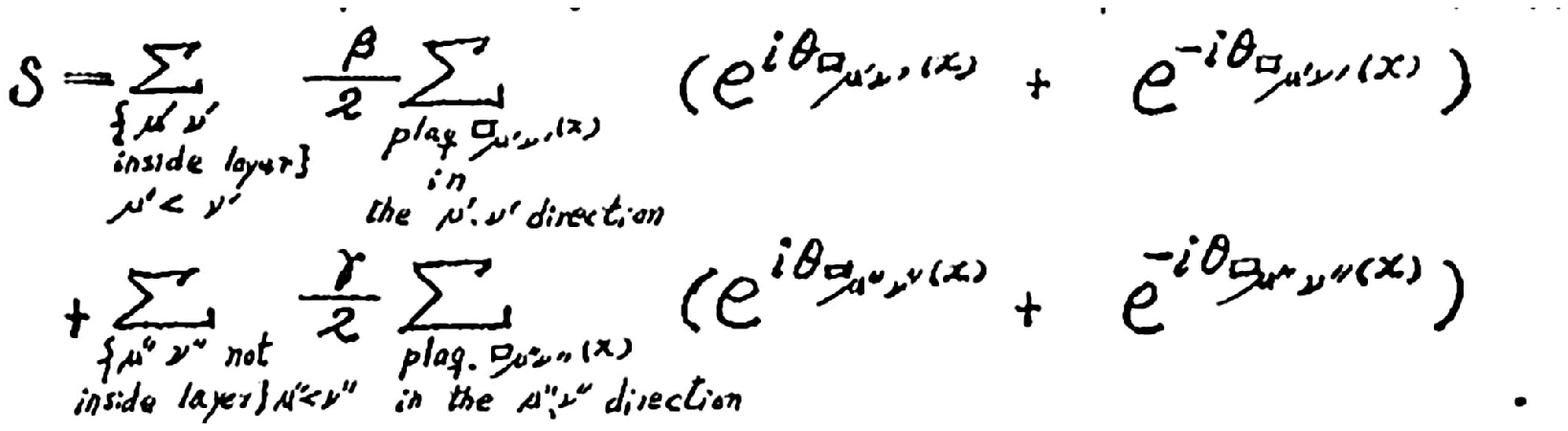}\\[1ex] 

where I quote handwritten formula from Nielsen and Fu \cite{Fu:1983ei} for historical reasons.
This action describes a sort of five-dimensional compact electrodynamics. As
we venture into new territory it is useful to establish a glossary:

\begin{itemize}
\item \textbf{pentaquum} is the ground state of the five-dimensional space
\item \textbf{fifth}, noun - fifth direction, or width/depth in that direction 
\item \textbf{pentark} is a fermi-like massive particle living in five dimensions
\item \textbf{pluon} is a counterpart of gluon, massless coloured particle
\item \textbf{penton} is a bound state of pentark and anti-pentark.
\end{itemize}

This model exhibits a rich phase structure, with the traditional Coulomb and
confined phases being joined by the Layer phase, which has following properties: 
        \begin{itemize}
        \item it exists for $D>5$, $d=D-1$, 
        \item but not for $D=5$ for non-Abelian theories
        \item in cQED massless ``photons'' propagate into fifth
        \item  yet propagators are supressed by $\gamma^{4N}$, where $N$ is
          the distance between layers and $\gamma$ is the coupling in the fifth dimension
        \item wilson loops follow the area law of the $d$-dimensional projection
        \item thus charged particles are confined to the layer, always
        \end{itemize}
   
 These results have been confirmed and extended in \cite{Gupta:1985he},
 \cite{Hulsebos:1994pa} and others. Unfortunately the case we find most
 compelling, the $SU(3)$, does not exhibit the desired behaviour. However, in
 \cite{Wang:2001ab} the authors use variational cumulant expansion to show the
 existence of the layer phase at finite temperature. In another work,
 \cite{Murata:2003rk} an interesting approach to the continuum limit of the
 five-dimensional gauge theory is suggested. But we  prefer to start from
 another viewpoint, and worry about the continuum limit later.

      \section{World Overview}
        
        \centerline{  \includegraphics[width=0.68\linewidth]{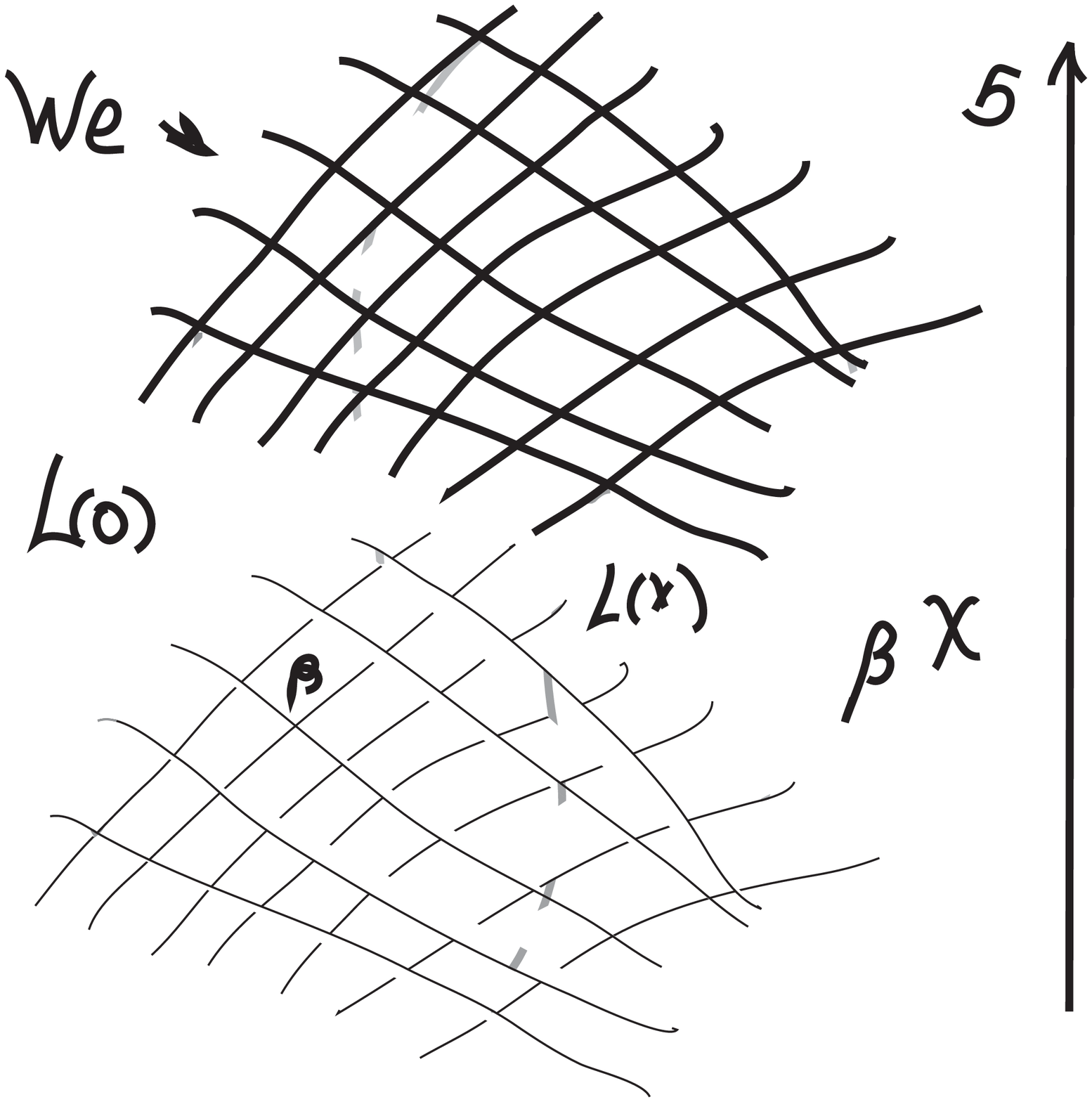}
        }
        
So let us consider what a desired theory should look like. It is       
      obviously $SU(3)$-symmetric gauge theory, which would mimic
   the four-dimensional gauge theory that QCD is. Now, philosophically speaking,  we should be able to see the
   effects of the fifth dimension from where we are. Otherwise we will be
   subject to the landscape
   problem which would make the model far less attractive. The Layer Phase in the
   original understanding does not allow such luxury, layers are
   self-contained and particle propagation is suppressed immensely. So the
   hope is that, when considered from another standpoint, five-dimensional
   non-isotropic theory will be not in the layer phase, yet has both required
   phenomenology of the real world and something else. So let us check if  ``traditional'' non-isotropic five dimensional gauge theory will do. 
Consider pure gauge theory on the lattice  with one coupling being smaller, the penta-coupling
 
         \[ S=-\frac{\beta}{N_{c}}\sum_{\mu\neq\nu=0..3}P_{\mu\nu}-\chi\frac{\beta}{N_{c}%
          }\sum_{\mu=0..3}P_{\mu4}%
          \]
where $P$ is the ordered product of the gauge fields along the plaquette.
Simulating it with the standard heat-bath on a penta-cubic lattice we can confirm
that, in the range of couplings, studied the Layer phase is not found. Usual
Coulomb and strong-coupling phases are present, and the plaquettes (square of the
field strength tensor)  change only
slowly with $\chi$, while Polyakov loops, the usual order parameter, do not go
anywhere near zero. 

While
it may sound disappointing, it actually is not. None of these variables is
truly physical in four-dimensional sense. Strictly speaking, the only physical
construction we can make with pure glue is Wilson loop / Polyakov loops
correlation, which (in four dimension)
corresponds to the propagation of a heavy meson. 
So let us define

   \[F_4(x)=\sum TrL_4(0)TrL_4(x) \;\;\;\;  F_5(x)=\sum Tr L_5(0)Tr L_5(x)\]

 which then describe the propagation of a particle-anti-particle bound state
    into the 4th and 5th dimension.

 \begin{figure}       
 \includegraphics[width=0.68\linewidth]{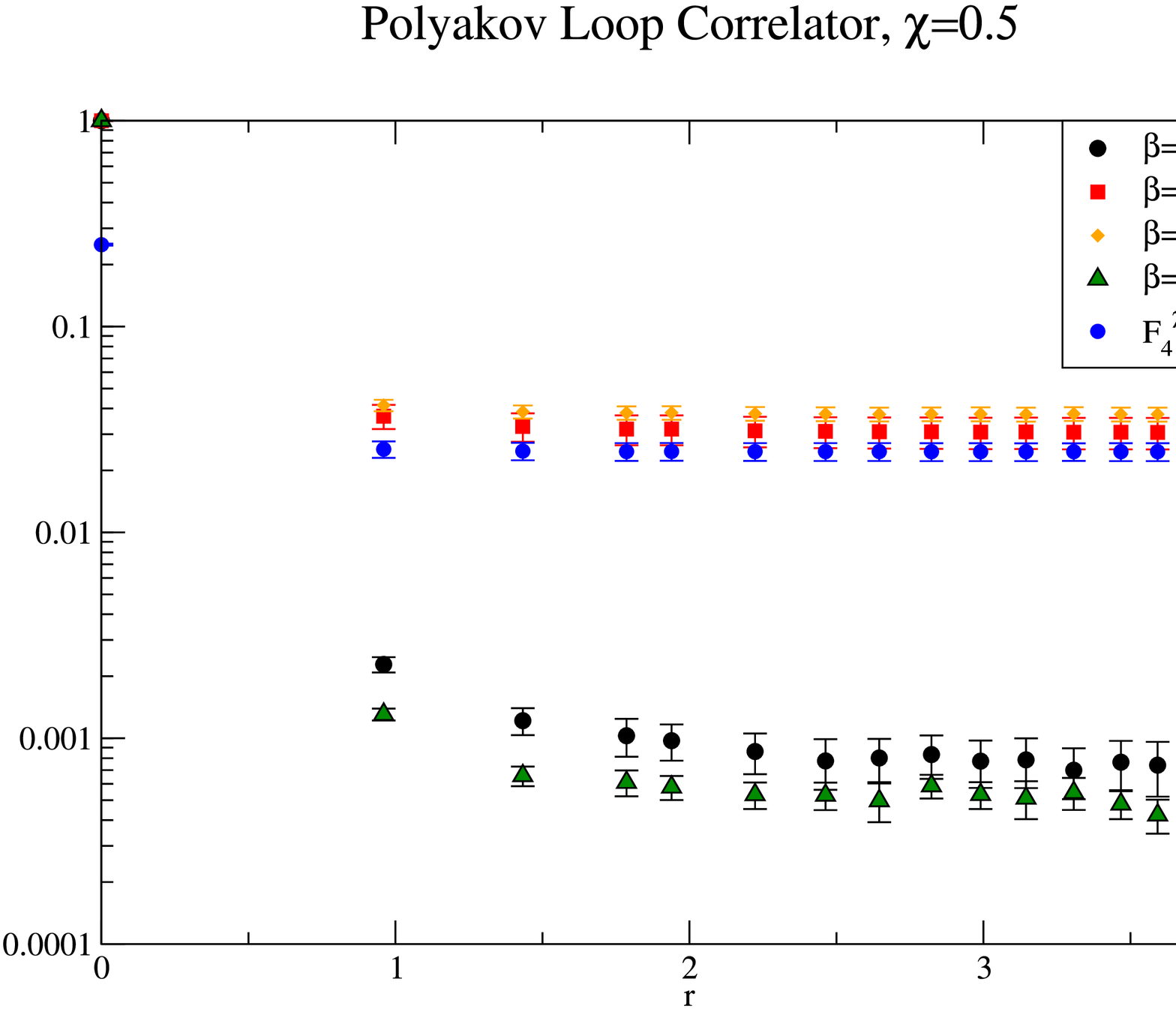}
\label{plot1}
\end{figure}

And here things start to look interesting. We are obviously in the Coulomb
phase. However, even at a very modest anisotropy
coefficient, we observe that fifth-correlator is supressed almost by two orders
of magnitude with respect to the four-correlator. The latter is reduced only
by a modest amount, roughly proprotionate to the naively expected
$\chi^2$. Moreover, the situation remains the same when changing the four
dimensional coupling towards the confined phase, while keeping the coupling ratio constant.   
    This indicates that dynamics of the gauge fields in pentaquum favours
    layerisation for the bound states while still granting free pass for the
    pentarks. Such situation is actually close to the desired theory, on one
    hand we have obviously four-dimensional layers for the pentons, which
    become mesons; and on the other hand, pentarks travel freely into fifth,
    giving us obvious explanation for the fact that we cannot observe
 free quarks. At the same time as propagation is not forbidden but just
 supressed, one may envision that with sufficient energy we can pick up a
 penton stuck to another layer and have it as virtual quark-anti-quark
 pair. This is far more intuitive than producing them out of vacuum.
     Of course this is more of a conjecture than a proof. Not only we are in a
     wrong phase for the four-dimensional world (which is known to be
     confining) but also somewhat away from the continuum limit. Moreover,
     these results are only applicable to heavy pentarks. Yet as we are still
     far from the connection to phenomenology, we do not know if our ``light''
     quarks are actually ``heavy'' in pentaquum.

      %%%%%%%%%%%%%%%%%%%%%%%%%%%%%%%%%%%%%%%%%%%%%%%%%%%%%%% 
 \section{Conclusion and Outlook} 
       
   We found a region in $5D$ space which is not exactly Nielsen
          Layer phase, but confines bound states to the layer, which is even
          more desirable and has chances to lead to observable predictions. It provides a
          compactification scenario alternative to Kaluza-Klein compact
          dimensions. It gives straightforward interpretation of confinement
          and virtual quark loops, which is qialitatively consistent with the
          traditional picture. 
Our intention now is to continue studying this model, calculate meson propagators and
confirm the conjecture that bound states get stuck to the layer.

Author is thankful to Holger Nielsen for letting him know about this model and
many useful discussions about  space and time and beyond.

\end{document}